\documentclass[journal]{IEEEtran}
\usepackage{cite}

\ifCLASSINFOpdf
  \usepackage[pdftex]{graphicx}
\else
\fi
\usepackage{amsmath}
\usepackage{amssymb}
\usepackage{bm}
\usepackage{algorithm,algorithmic}
\newcommand{\tr}{\mathrm{tr}}
\usepackage{tabularx,booktabs}

\newcommand{\squeezeup}{\vspace{-1mm}}

\begin{document}
%
\title{Distributed Dimension Reduction for Distributed Massive MIMO C-RAN with Finite Fronthaul Capacity}
%
%
%

\author{\IEEEauthorblockN{Fred Wiffen\IEEEauthorrefmark{1}\IEEEauthorrefmark{2},
Woon Hau Chin\IEEEauthorrefmark{1}\IEEEauthorrefmark{3} and
Angela Doufexi\IEEEauthorrefmark{1}} \\
\IEEEauthorblockA{\IEEEauthorrefmark{1}Department of Electrical \& Electronic Engineering, University of Bristol, U.K.} \\
\IEEEauthorblockA{\IEEEauthorrefmark{2}Bristol Research \& Innovation Laboratory, Toshiba Europe Limited, U.K.} \\
\IEEEauthorblockA{\IEEEauthorrefmark{3}Viavi Solutions, Stevenage, U.K.} \\
\IEEEauthorblockA{Email: fred.wiffen@bristol.ac.uk}}

\maketitle

\begin{abstract}
The use of a large excess of service antennas brings a variety of performance benefits to distributed MIMO C-RAN, but the corresponding high fronthaul data loads can be problematic in practical systems with limited fronthaul capacity. In this work we propose the use of lossy dimension reduction, applied locally at each remote radio head (RRH), to reduce this fronthaul traffic. We first consider the uplink, and the case where each RRH applies a linear dimension reduction filter to its multi-antenna received signal vector. It is shown that under a joint mutual information criteria, the optimal dimension reduction filters are given by a variant of the conditional Karhunen-Loeve transform, with a stationary point found using block co-ordinate ascent. These filters are then modified such that each RRH can calculate its own dimension reduction filter in a decentralised manner, using knowledge only of its own instantaneous channel and network slow fading coefficients. We then show that in TDD systems these dimension reduction filters can be re-used as part of a two-stage reduced dimension downlink precoding scheme. Analysis and numerical results demonstrate that the proposed approach can significantly reduce both uplink and downlink fronthaul traffic whilst incurring very little loss in MIMO performance.
\end{abstract}


%
\IEEEpeerreviewmaketitle

\section{Introduction}

%
%
%
%
\IEEEPARstart{D}{istributed} multi-user MIMO systems use multiple geographically distributed remote radio head (RRHs) to co-operatively serve multiple users within an extended coverage area. This distribution of the base station (BS) antennas provides macro-diversity to reduce variations in pathloss, improving uniformity of service compared to centralised MIMO configurations. It has seen increased interest in recent years due to the growing popularity of the cloud radio access network (C-RAN) architecture -- where multiple RRHs are connected via fronthaul links to a shared central processor (CP), enabling joint MIMO processing of all signals.

Here, we consider a C-RAN system where $L$ RRHs each equipped with $M$ antennas jointly serve $K$ single antenna users. It is well known that operating such a system in the `massive' regime, by deploying a large overall excess of BS antennas, $ML \gg K$, can bring significant benefits -- providing array gain, diversity against both fast and slow fading, and improving the channel condition such that low complexity, scalable linear processing techniques are near-optimal. 

However, the use of a large number of antennas brings a proportionate increase in the quantity of data that must be transferred over the fronthaul connections. This can present a challenge for practical systems which may rely on limited capacity fronthaul links based on wireless point-to-point, ethernet, or shared fibre \cite{7456186}.

In this work we consider the use of lossy distributed dimension reduction to reduce the amount of fronthaul data in distributed massive MIMO C-RAN systems. Whilst there has been interest in distributed dimension reduction techniques within the broader literature, its explicit application to distributed MIMO systems has yet to be studied in detail. 

We propose to apply linear dimension reduction at each RRH to produce a low dimension effective MIMO system that exploits the inherent sparsity of massive MIMO to approximate the performance of the large-scale distributed MIMO system whilst reducing the quantity of data that must be transferred over the fronthaul connections. We show that this approach can be employed on both the uplink \& downlink, and constitutes a bidirectional approach to fronthaul load reduction that is well suited to time division duplex (TDD) systems.

\subsection{Related Work} 
Dimension reduction plays a role in many data compression/reduction schemes, and exploits the fact that many high dimension signals have a sparse representation in an alternative signal basis and can therefore be represented, perfectly or approximately, using a reduced number of coefficients.

In \cite{4016296} a number of problems related to finding reduced dimension distributed representations of an observed signal vector are investigated. It is shown that when linear dimension reduction is applied locally at a single node, the MMSE-optimal dimension reduction filter is given by the principal eigenvectors of the conditional covariance matrix at that node. They call this the `conditional' Karhunen-Loeve transform due to its connection to the well known Karhunen-Loeve transform (the eigenvectors of the marginal covariance matrix \cite{952802}). 

The work in \cite{4276987} addresses the related problem of finding the reduced dimension signals from which some other correlated quantity is linearly estimated. The MMSE-optimal filters are shown to depend on various second order signal statistics, with a stationary point found using block coordinate ascent.

Some data reduction strategies using forms of dimension reduction have previously been proposed for distributed MIMO C-RAN systems. On the uplink, a simple dimension reduction strategy is proposed for reducing fronthaul data in \cite{8337794}, where each RRH takes simple unweighted sums of the signals received at different antennas in different time/frequency slots (i.e. binary dimension reduction filters). In our previous work in \cite{9120669}, we propose an uplink scheme that achieves dimension reduction at each RRH by matched filtering the received signal using a subset of the local user channel vectors. The channel vector subsets for all RRHs are selected jointly at the CP using full CSI to achieve a good overall signal representation. 

On the downlink, the popular sparse beamforming approach first outlined in \cite{6920005} can be seen as a form of dimension reduction for fronthaul data reduction. Dimension reduction is also applied at a network level in \cite{7880689}, where subsets of RRHs are deactivated to improve network energy efficiency.

Despite its clear applicability, there has been little direct investigations into the optimisation of distributed dimension reduction for the distributed MIMO C-RAN context. Our work here seeks to begin to address this gap in the literature.

\subsection{Paper Structure}
The paper is structured as followed: Section \ref{sec:SystemModel} defines the system model for the distributed MIMO C-RAN uplink and downlink, before Section \ref{sec:UplinkDistributedDimensionReduction} provides an introductory discussion on the use of distributed dimension reduction on the distributed massive MIMO C-RAN uplink. Section \ref{sec:SumRateCriteria} addresses \& analyses the problem of jointly optimising the dimension reduction filters under a joint mutual information criteria. An alternative fully decentralised approach in which each RRH calculates its own dimension reduction filter using partial CSI is proposed in Section \ref{sec:FullyDecentralised}, before Section \ref{sec:PracticalAspects} discusses some practical aspects of distributed dimension reduction. Section \ref{sec:DownlinkDimensionReduction} extends the dimension reduction concept to the distributed MIMO C-RAN downlink. Finally, Section \ref{sec:NumericalResults} provides numerical results demonstrating the performance of the proposed approaches.

\subsection{Notation}
We use the notations $x$, $\mathbf{x}$ and $\mathbf{X}$ to represent scalar, vector and matrix quantities respectively. The conjugate, transpose and conjugate transpose of $\mathbf{X}$ are represented $\mathbf{X}^*$, $\mathbf{X}^T$ and $\mathbf{X}^\dagger$. $\big[\mathbf{X}\big]_{i,j}$ represents the entry of $\mathbf{X}$ at row $i$, column $j$. Italics, $\bm{X}$, are used in places to emphasise where a quantity is treated as a random variable. Trace, determinant, 2-norm and expectation are denoted $\tr(.)$, $\det(.)$, $\Vert . \Vert$ and $\mathbb{E}\{.\}$.

\section{System Model}
\label{sec:SystemModel}
We consider a C-RAN system in which $K$ single antenna users are co-operatively served by $L$ geographically distributed multi-antenna RRHs, each equipped with $M$ antennas. We focus specifically on massive MIMO C-RAN systems, in which the total number of base station antennas is significantly larger than the number of users being served, $ML \gg K$. 

The RRHs are connected to a single shared central processor (CP) via fronthaul connections, and uplink symbol detection \& downlink symbol precoding are performed at the CP using signals from/for all RRHs.

We assume that the network operates in time division duplex mode, such that the uplink and downlink channels are reciprocal and can be estimated using uplink pilots transmitted by the users. A detailed discussion on CSI requirements is provided in Section \ref{sec:PracticalAspects}. All numerical results provided in this work use the configurations described in Section \ref{sec:NumericalResults}.

\subsection{Distributed MIMO Uplink}
The uplink signal received locally at RRH $l$, $\mathbf{y}_l \in \mathbb{C}^M$ is
\begin{equation}
	\mathbf{y}_l = \sqrt{\rho}\mathbf{H}_l\mathbf{x} + \bm{\eta}	_l,
\end{equation}
where $\rho$ is the user SNR, $\mathbf{x} = \begin{bmatrix} x_1 & x_2 & \ldots & x_K \end{bmatrix}^T$ contains the $K$ unit variance transmit symbols, $x_k \sim \mathcal{CN}(0,1)$, and $\bm{\eta}_l \sim \mathcal{CN}(0,\mathbf{I}_M)$ is receiver noise. The uplink channel matrix, $\mathbf{H}_l \in \mathbb{C}^{M \times K}$, is structured
\begin{equation}
	\mathbf{H}_l = \begin{bmatrix} \sqrt{p_1}\mathbf{h}_{l,1} & \ldots & \sqrt{p_K}\mathbf{h}_{l,K} 	
 \end{bmatrix} = \bar{\mathbf{H}}_l\mathbf{P}^{1/2}
\end{equation}
where $\mathbf{h}_{l,k} \in \mathbb{C}^M$ is the channel between user $k$ and RRH $l$, and $p_k = [\mathbf{P}]_{k,k}$ the power control coefficient for user $k$ (absorbed into the channel matrix to simplify notation). It is assumed that each $\mathbf{H}_l$ is full rank, i.e. $r = \mathrm{rank}(\mathbf{H}_l) = \min(M,K)$, and that,
\begin{equation}
	 \frac{1}{M}\mathbb{E}\{\Vert \mathbf{h}_{l,k}\Vert^2\} = \beta_{l,k},	
\end{equation}
where $\beta_{l,k}$ is the pathloss (slow fading) coefficient for the channel between user $k$ and RRH $l$.

\subsection{Distributed MIMO Downlink}
For the downlink we consider a TDD configuration where the downlink channel is reciprocal to the uplink channel and the received signal at user $k$ is
\begin{equation}
	y_k = \sum_{l=1}^L\mathbf{h}_{l,k}^\dagger\mathbf{s}_l + \eta,	
\end{equation}
where $\mathbf{s}_l \in \mathbb{C}^M$ is the precoded signal transmitted by RRH $l$ and $\eta \sim \mathcal{CN}(0,1)$ receiver noise. This can be written more compactly in vector form
\begin{equation}
	\mathbf{y} = \sum_{l=1}^K \bar{\mathbf{H}}_l^\dagger\mathbf{s}_l + \bm{\eta}.
\end{equation}
We assume that the transmissions from each RRH are subject to individual \& identical power constraints,
\begin{align}
\label{eq:powerconstraint}
	\mathbb{E}\{\mathbf{s}_l^\dagger\mathbf{s}_l\} \leq P.	
\end{align}

\section{Uplink Distributed Dimension Reduction}
\label{sec:UplinkDistributedDimensionReduction}
\begin{figure}[!t]
\centering
\includegraphics[trim={8cm 9.5cm 4cm 2cm},clip,width=0.95\linewidth]{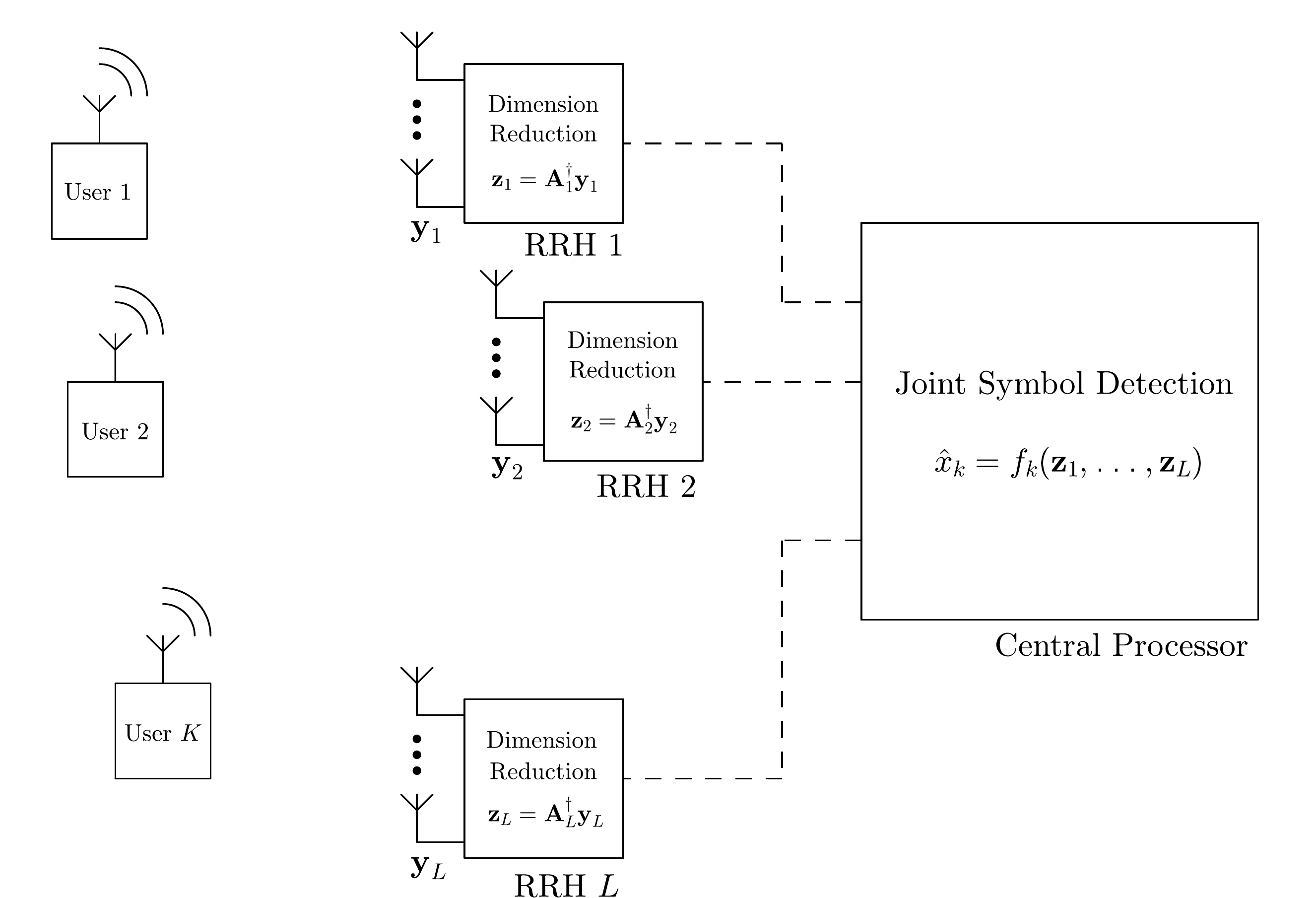}
\caption{Dimension reduction for uplink distributed massive MIMO C-RAN.}
\label{fig:DimRed}
\end{figure}

We begin by considering the distributed MIMO uplink. In our proposed scheme, shown in Figure \ref{fig:DimRed}, each multi-antenna RRH in the network receives an $M$-dimension signal,
\begin{equation}
	\mathbf{y}_l = \sqrt{\rho}\mathbf{H}_l\mathbf{x} + \bm{\eta}_l,
\end{equation}
and applies a linear dimension reduction filter, $\mathbf{A}_l \in \mathbb{C}^{M \times N}$, to produce an $N$-dimension signal,
\begin{equation}
	\mathbf{z}_l = \mathbf{A}_l^\dagger\mathbf{y}_l,
\end{equation}
which is then transferred over fronthaul to the CP for joint detection of the user transmit symbols, 
\begin{equation}
	\hat{x}_k = f_k(\mathbf{z}_1,\ldots,\mathbf{z}_L).
\end{equation}
Without loss of generality we may restrict our attention to semi-orthogonal filters\footnote{Using the QR decomposition, an arbitrary filter, $\tilde{\mathbf{A}}_l \in \mathbb{C}^{M \times N}$, with rank $N$ can be represented, $\tilde{\mathbf{A}}_l  = \mathbf{A}_l\mathbf{R}_l$, where $\mathbf{A}_l \in \mathbb{C}^{M \times N}$ is semi-orthogonal and $\mathbf{R}_l \in \mathbb{C}^{N \times N}$ is invertible. Since $\mathbf{R}_l$ is invertible it does not affect the information content of $\mathbf{z}_l = \tilde{\mathbf{A}}_l^\dagger\mathbf{y}_l$.}, i.e where
\begin{equation}
	\mathbf{A}_l^\dagger\mathbf{A}_l = \mathbf{I}_N.
\end{equation}
The reduced dimension signal produced at each RRH can then be treated as the output of an equivalent reduced dimension MIMO channel,
\begin{equation}
	\mathbf{z}_l = \sqrt{\rho}\mathbf{H}_l^{\scriptscriptstyle \mathrm{RD}}\mathbf{x} + \bm{\omega}_l
\end{equation}
where $\mathbf{H}_l^{\scriptscriptstyle \mathrm{RD}} = \mathbf{A}_l^\dagger\mathbf{H}_l$ and $\bm{\omega}_l \sim \mathcal{CN}(0,\mathbf{I}_N)$.

Under MMSE detection, the uplink rate of user $k$ is then
\begin{equation}
	\mathcal{R}_k^{\scriptscriptstyle \mathrm{UL}} = \log_2\big(1 + \mathrm{SINR}_k\big)
\end{equation}
where
\begin{equation}
	\mathrm{SINR}_k = \frac{1}{\Big[\Big(\mathbf{I}_K + \rho\sum_{l=1}^L\mathbf{H}_l^{\scriptscriptstyle \mathrm{RD}\dagger}\mathbf{H}_l^{\scriptscriptstyle \mathrm{RD}}\Big)^{-1}\Big]_{k,k}	} - 1.
\end{equation}

From fundamental linear algebra, the information-carrying component of $\mathbf{y}_l$ lies in a $r = \mathrm{min}(M,K)$-dimension vector space, and hence a minimum of $N = r$ coefficients per RRH are required for a lossless signal representation. Obtaining a perfect representation of all received signals at the CP therefore requires a minimum of $rL = \mathrm{min}(ML,KL)$ coefficients to be transferred over fronthaul. This can lead to large fronthaul loads in large scale systems where $ML \gg K$.

We focus instead on the \textit{lossy} case where $N < r$, and aim to exploit the inherent \textit{joint sparsity} in the distributed massive MIMO signals to create a reduced dimension system that approximates the performance of the full dimension system.

At a basic level, we note that providing the overall signal dimension is greater than the number of users, $NL \geq K$, then randomly chosen dimension reduction filters are generally sufficient to produce an overall reduced dimension channel, $\mathbf{H}_{\scriptscriptstyle \mathrm{G}}$, with full rank, and hence achieve the full MIMO multiplexing gain of $K$. However, random filters do not capture the diversity gains provided by a system with a large overall excess of antennas, and can result in a poorly conditioned system\footnote{For the independent Rayleigh fading case it can be shown that using random dimension filters is equivalent to simply reducing the number of antennas at each RRH from $M$ to $N$.}


On the other hand, it is clear that if the dimension reduction filters are optimised with respect to the channel, the reduced dimension MIMO system should be able to outperform a smaller distributed MIMO system with only $N$ antennas deployed at each RRH -- since this can be modelled as a special case of linear dimension reduction where $\mathbf{A}_l$ contains the first $N$ columns of $\mathbf{I}_M$. For example, we could deploy $M$ antennas and choose $\mathbf{A}_l$ to be the `best' $N$ columns of $\mathbf{I}_M$. This is the antenna selection method that has been widely studied for centralised MIMO systems and is known to be able to achieve a significant proportion of the system capacity using a reduced number of antennas \cite{1341263}. The work here can be seen as a generalisation of this for distributed systems, where $\mathbf{A}_l$ is allowed to take any value.

\squeezeup
\section{Filter Design under Joint Mutual Information Criteria}
\label{sec:SumRateCriteria}
Here, we consider the problem of finding the dimension reduction filters that maximise the joint mutual information between the reduced dimension signals and the user symbols,
\begin{align}
\label{eq:sumrateopt}
	&\underset{\mathbf{A}_1,\ldots,\mathbf{A}_L}{\mathrm{maximise}} \quad \mathcal{I}\big(\mathbf{z}_1,\ldots,\mathbf{z}_L;\mathbf{x}\big) \nonumber \\
	&\mathrm{subject \ to} \quad \mathbf{A}_l^\dagger\mathbf{A}_l = \mathbf{I}_N. 
\end{align} 
This is equivalent to finding the dimension reduction filters that maximise the total amount of information that the set of reduced dimension signals \textit{jointly} capture about the transmitted symbols, given by
\begin{equation*}
	\mathcal{I}\big(\mathbf{z}_1,\ldots,\mathbf{z}_L;\mathbf{x}\big) = \log_2\det\big(\mathbf{I}_K + \rho\sum_{l=1}^L\mathbf{H}_l^\dagger\mathbf{A}_l\mathbf{A}_l^\dagger\mathbf{H}_l\big),
\end{equation*}
It also represents the maximum sum rate of the reduced dimension MIMO system under optimal joint symbol detection.

Unfortunately \eqref{eq:sumrateopt} is non-convex in the $\mathbf{A}_l$, making find a global maximum challenging. Here, we instead show that a stationary point can be found using block coordinate ascent. We begin by considering the problem of finding the optimal dimension reduction filter, $\mathbf{A}_l$ at RRH $l$, under the assumption that the other dimension reduction filters, $\mathbf{A}_1,\ldots,\mathbf{A}_{l-1},\mathbf{A}_{l+1},\ldots,\mathbf{A}_L$, are fixed.

\subsection{Optimisation at Single RRH}
Using the chain rule, the joint mutual information can be expanded
\begin{equation}
\label{eq:chainrule}
	\mathcal{I}\big(\mathbf{z}_1,\ldots,\mathbf{z}_L;\mathbf{x}\big) = \mathcal{I}\big(\mathbf{z}_l^\mathsf{c};\mathbf{x}\big)  + \mathcal{I}\big(\mathbf{z}_l;\mathbf{x}\big\vert \mathbf{z}_l^\mathsf{c}\big) 
\end{equation}
where $\mathbf{z}_l^\mathsf{c} = \{\mathbf{z}_1,\ldots,\mathbf{z}_{l-1},\mathbf{z}_{l+1},\ldots,\mathbf{z}_L \}$. In this expansion the first term, $\mathcal{I}\big(\mathbf{z}_l^\mathsf{c};\mathbf{x}\big)$, represents the joint mutual information when the CP has access to the reduced dimension signals from the other $L-1$ RRHs. The second term, $\mathcal{I}\big(\mathbf{z}_l;\mathbf{x}\big\vert \mathbf{z}_l^\mathsf{c}\big)$, represents the \textit{conditional} information contribution of RRH $l$ -- i.e. the increase when the CP also has access to $\mathbf{z}_l$. Applying the well-known matrix determinant lemma, this can be shown to be
\begin{equation}
	\mathcal{I}\big(\mathbf{z}_l;\mathbf{x}\big\vert \mathbf{z}_l^\mathsf{c}\big) = \log_2\det\Big(\mathbf{I}_N + \rho\mathbf{A}_l^\dagger\mathbf{H}_l\mathbf{Q}_l\mathbf{H}_l^\dagger\mathbf{A}_l\Big),
\end{equation}
where
\begin{equation}
	\mathbf{Q}_l = \Big(\mathbf{I}_K + \rho\sum_{j \neq l}\mathbf{H}_j^\dagger\mathbf{A}_j\mathbf{A}_j^\dagger\mathbf{H}_j\Big)^{-1}.
\end{equation}
Since the first term in \eqref{eq:chainrule} does not depend on $\mathbf{A}_l$, the optimal dimension reduction filter for RRH $l$ is the solution to
\begin{align}
\label{eq:max_cond}
	&\underset{\mathbf{A}_l}{\mathrm{maximise}} \quad \log_2\det\Big(\mathbf{I}_N + \rho\mathbf{A}_l^\dagger\mathbf{H}_l\mathbf{Q}_l\mathbf{H}_l^\dagger\mathbf{A}_l\Big) \nonumber \\	
	&\mathrm{subject \ to} \quad \mathbf{A}_l^\dagger\mathbf{A}_l = \mathbf{I}_N 
\end{align}
By the Poincare separation theorem \cite{bellman1997introduction}, this has a single global maximum that is achieved when the columns of $\mathbf{A}_l$ are the eigenvectors corresponding to the $N$ largest eigenvalues of $\mathbf{H}_l\mathbf{Q}_l\mathbf{H}_l^\dagger$,
\begin{equation}
\label{eq:princ_eig}
	\mathbf{A}_l = N \ \mathrm{principal \ eigenvectors \ of} \ \mathbf{H}_l\mathbf{Q}_l\mathbf{H}_l^\dagger.
\end{equation}

The optimal dimension reduction filter at RRH $l$ contains the $N$ principal eigenvectors of the conditional covariance matrix $\mathbb{E}\big\{\mathbf{y}_l\mathbf{y}_l^\dagger\big\vert \mathbf{z}_l^\mathsf{c}\big\} = \mathbf{I}_M + \rho\mathbf{H}_l\mathbf{Q}_l\mathbf{H}_l^\dagger$, conditioned with respect to the signals supplied by the other RRHs. This is an instance of the \textit{conditional Karhunen Loeve transform} (CKLT) as previously defined in \cite{4016296}. For the special case of a single RRH ($L=1$) we have $\mathbf{Q}_l = \mathbf{I}_K$, and the dimension reduction filter is the Karhunen-Loeve transform (KLT) -- the filter that also maximises signal power, $\mathbb{E}\{ \Vert \mathbf{z}_l \Vert^2 \}$.
\squeezeup
\subsection{Block Coordinate Ascent}
\label{sec:BCA}

Since the CKLT achieves the global maximum to \eqref{eq:max_cond}, the filters can be jointly optimised using block coordinate ascent by successively updating the CKLT at each RRH in turn. Since the objective function \eqref{eq:sumrateopt} is bounded and increases monotonically with each update, convergence to a stationary point is assured. In our simulations here we initialise the coordinate ascent using the KLT filters.


\squeezeup
\subsection{Insights}
\begin{figure}[!t]
\centering
\includegraphics[width=3.6in]{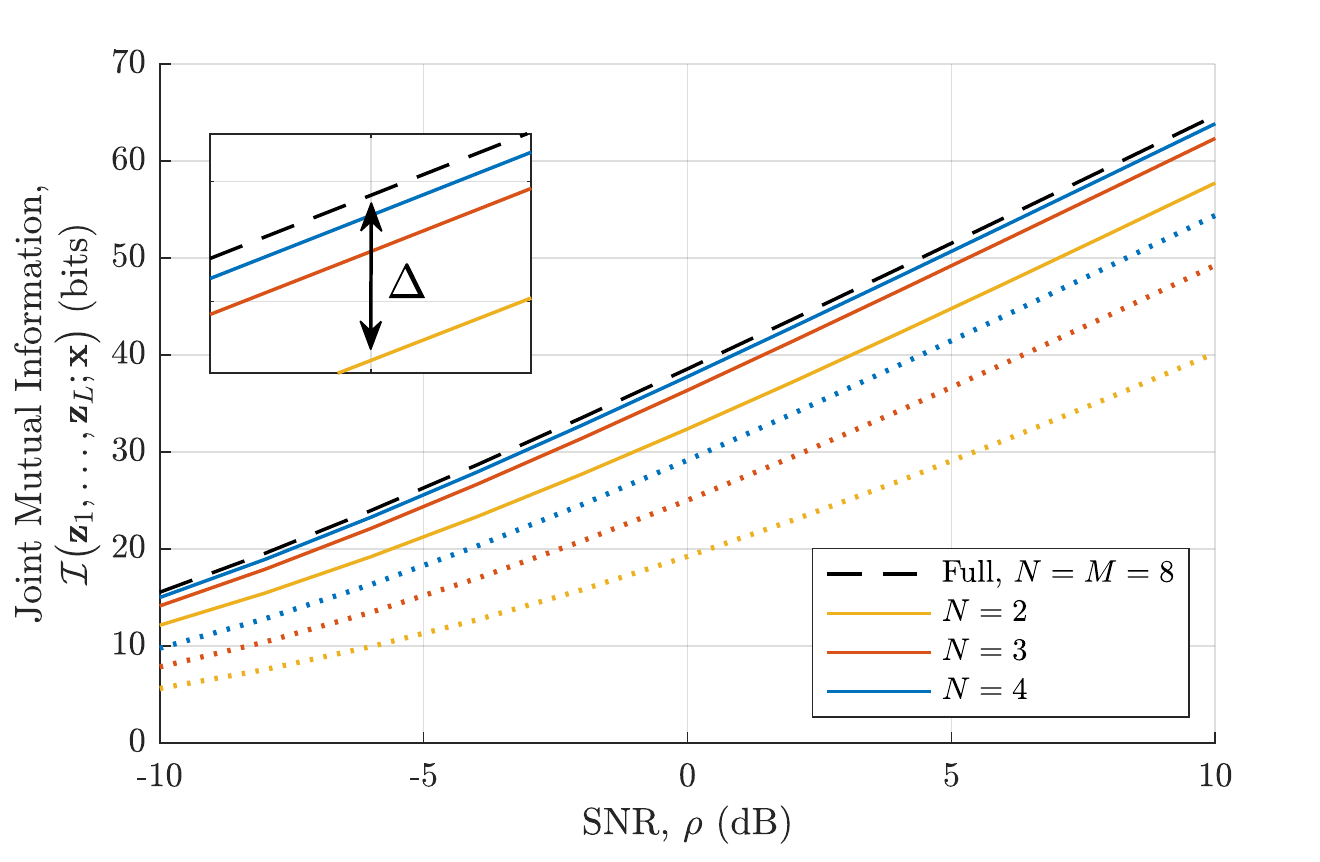}
\caption{Joint mutual information scaling with SNR under dimension reduction, $L=4,K=8,M=8,\rho = 5$ dB. Solid lines: centralised dimension reduction. Dotted lines: reduced number of antennas per RRH ($M' = N$).}
\label{fig:sum_rate}
\end{figure}

The loss in information due to dimension reduction is
\begin{align}
	\Delta &= \mathcal{I}\big(\mathbf{y}_1,\ldots,\mathbf{y}_L;\mathbf{x}\big) - \mathcal{I}\big(\mathbf{z}_1,\ldots,\mathbf{z}_L;\mathbf{x}\big) \\
	&= \log_2\frac{\det\big(\mathbf{I}_K + \rho\sum_{l=1}^L\mathbf{H}_l^\dagger\mathbf{H}_l\big)}{\det\big(\mathbf{I}_K + \rho\sum_{l=1}^L\mathbf{H}_l^{\scriptscriptstyle \mathrm{RD}\dagger}\mathbf{H}_j^{\scriptscriptstyle \mathrm{RD}}\big)}.
\end{align}
Providing the reduced dimension system is full rank, the overall loss is bounded at high SNR,
\begin{equation}
	\lim_{\rho \to \infty} \Delta = \log_2\frac{\det\big(\sum_{l=1}^L\mathbf{H}_l^\dagger\mathbf{H}_l\big)}{\det\big(\sum_{l=1}^L\mathbf{H}_l^{\scriptscriptstyle \mathrm{RD}\dagger}\mathbf{H}_l^{\scriptscriptstyle \mathrm{RD}}\big)},
\end{equation}
and represents a vanishing fraction of the joint mutual information captured by the reduced dimensions signals. This can be seen  in Figure \ref{fig:sum_rate}, for a system with $L=4,M=8,K=8$. Note that with $N \geq 3$, the overall loss in information due to applying dimension reduction is very small. There is a clear benefit to using a larger number of antennas at each RRH and then applying dimension reduction compared to simply deploying a reduced number of antennas.

\squeezeup
\section{Fully Decentralised Dimension Reduction using Partial CSI}
\label{sec:FullyDecentralised}
The method in Section \ref{sec:SumRateCriteria} for jointly optimising the uplink dimension reduction filters requires full network CSI, and therefore in practice must be implemented centrally at the CP. We now investigate an alternative decentralised approach, where each RRH locally calculates its own uplink filter using full knowledge of its own channel matrix, $\mathbf{H}_l$, whilst treating the channel matrices for the other $L-1$ RRHs as unknown random quantities, $\bm{H}_j$. 

We can adapt the joint mutual information maximisation method from Section \ref{sec:SumRateCriteria} by choosing $\mathbf{A}_l$ to instead maximise the expected value of the quantity
\begin{align}
\label{eq:decentral_opt}
	&\underset{\mathbf{A}_l}{\mathrm{maximise}} \quad \mathbb{E}\Big\{\log_2\det\Big(\mathbf{I}_N + \rho\mathbf{A}_l^\dagger\mathbf{H}_l\bm{Q}_l\mathbf{H}_l^\dagger\mathbf{A}_l\Big)\Big\} \nonumber \\	
	&\mathrm{subject \ to} \quad \mathbf{A}_l^\dagger\mathbf{A}_l = \mathbf{I}_N, 
\end{align}
where $\bm{Q}_l$ is now a random variable that depends on the random channel and filter realisations at the other RRHs
\begin{equation}
	\bm{Q}_l = \Big(\mathbf{I}_K + \rho\sum_{j \neq l}\bm{H}_j^\dagger\bm{A}_j\bm{A}_j^\dagger\bm{H}_j\Big)^{-1}.
\end{equation}
In this expression the optimal $\mathbf{A}_l$ depends on both the other $\bm{A}_j$ and the probability distributions for the $\bm{H}_j$, making finding a solution challenging. We can significantly simplify this by instead choosing $\mathbf{A}_l$ to maximise a lower bound for \eqref{eq:decentral_opt}, shown in \eqref{eq:lower_bound}. This lower bound is established in two stages:
\begin{itemize}
\item First, we use the lower bound in \eqref{eq:Jensen} to eliminate the dependence on the dimension reduction filters at the other RRHs, by noting that $\bm{A}_j\bm{A}_j^\dagger$ is an orthogonal projection. This is equivalent to designing the dimension reduction filter at RRH $l$ under the condition that the other $L-1$ RRHs each supply their full signal to the CP.
\item Second, we exploit the convexity of \eqref{eq:Jensen} \cite{DBLP:journals/corr/Kim15e} and apply Jensen's inequality to produce an objective function that depends only on the channel statistics $\mathbb{E}\big\{\bm{H}_j^\dagger\bm{H}_j\big\}$.
\end{itemize}

\begin{figure*}[!t]
\normalsize
\begin{align}
    \mathbb{E}\Big\{\log_2\det\Big(\mathbf{I}_N + \rho\mathbf{A}_l^\dagger\mathbf{H}_l\bm{Q}_l\mathbf{H}_l^\dagger\mathbf{A}_l\Big)\Big\} &\geq \mathbb{E}\Big\{\log_2\det\Big(\mathbf{I}_N + \rho\mathbf{A}_l^\dagger\mathbf{H}_l\Big(\mathbf{I}_K + \rho\sum_{j \neq l}\bm{H}_j^\dagger\bm{H}_j\Big)^{-1}\mathbf{H}_l^\dagger\mathbf{A}_l\Big)\Big\}.
\label{eq:Jensen} \\
    &\geq \log_2\det\Big(\mathbf{I}_N + \rho\mathbf{A}_l^\dagger\mathbf{H}_l\Big(\mathbf{I}_K + \rho\sum_{j \neq l}\mathbb{E}\big\{\bm{H}_j^\dagger\bm{H}_j\big\}\Big)^{-1}\mathbf{H}_l^\dagger\mathbf{A}_l\Big).
\label{eq:lower_bound}
\end{align}
\hrulefill
\vspace*{4pt}
\end{figure*}

Under the mild assumption that the different user channel realisations vary independently, $\mathbb{E}\{\mathbf{h}_{l,k}^\dagger\mathbf{h}_{l,j}\} = 0$, $\mathbb{E}\{\bm{H}_j^\dagger\bm{H}_j\} = \mathbf{\Psi}_j$ is a diagonal matrix that depends only on the user power control and pathloss coefficients,
\begin{equation}
	\big[\mathbf{\Psi}_j\big]_{k,k} = p_k\mathbb{E}\{\Vert \mathbf{h}_{j,k}\Vert^2\} = p_k\beta_{j,k}M
\end{equation}
The dimension reduction filters can then be calculated as
\begin{equation}
	\mathbf{A}_l = N \mathrm{\ princ. \ eigenv. \ of \ } \mathbf{H}_l\big(\mathbf{I}_K + \rho\sum_{j \neq l}\mathbf{\Psi}_j\big)^{-1}\mathbf{H}_l^\dagger.
\end{equation}

We refer to this filter as the D-CKLT. In contrast with the CKLT, with the D-CKLT dependencies between RRHs are accounted for using only the slow fading characteristics. This seems an intuitively reasonable strategy for a distributed MIMO network, where the $\beta_{l,k}$ may vary by orders of magnitude due to the physical distribution of users and RRHs.


After each RRH calculates and applies its dimension reduction filters, the CP jointly detects the user symbols using full network reduced dimension CSI. Thus, the use of partial CSI impacts the system performance only through the choice of dimension reduction filter. Numerical results presented in Section \ref{sec:NumericalResults} show that the use of decentralised filter design often incurs only a small performance sacrifice. 

\squeezeup
\section{Practical Aspects}
\label{sec:PracticalAspects}
In practical mobile channels with finite channel coherence time the dimension reduction filters must be regularly updated, and hence the computational and signalling overheads associated with distributed dimension reduction must be considered.
\squeezeup
\subsection{Computational Complexity}
\label{sec:comp_comp}
The centralised approaches to filter design are iterative in nature, with each iteration requiring a $K \times K$ matrix inversion with complexity $\mathcal{O}(K^3)$, an $M \times M$ eigendecomposition, complexity $\mathcal{O}(M^3)$, and some matrix multiplications. Simulations indicate that the block coordinate procedure typically converges with a small number ($\sim$3) of updates at each RRH.


In contrast, the decentralised approach has a closed form solution, requiring a single $M \times K$ singular value decomposition with complexity $\mathcal{O}(MK^2)$ per RRH. This computation is performed locally at the RRHs, meaning the decentralised approach requires the RRHs to have enhanced computational capabilities compared to the centralised approach.
\squeezeup
\subsection{Fronthaul CSI Overheads}
In the centralised approach, the CP requires full network CSI for calculating the dimension reduction filters, which must then be transferred back to the RRHs. Assuming CSI is initially obtained at the RRHs through uplink pilot symbols, additional data must be transferred both to and from the CP.

One benefit of the decentralised approach is the reduction in overheads, since only the reduced dimension MIMO channels, $\mathbf{H}_l^{\scriptscriptstyle \mathrm{RD}}$, need to be transferred from RRHs to CP (with only changes in large scale fading communicated back). The number of coefficients that must be transferred, per channel coherence block, for each scheme are shown in Table I.

\begin{table}

\renewcommand{\arraystretch}{1.4}
\footnotesize
\caption{CSI-related Fronthaul Data Overheads}
\centering
\begin{tabular}{l | c | c}
 \hline Method & RRH to CP & CP to RRH \\ \hline 
  Standard Distributed MIMO & $\mathbf{H}_l$, $MK$ coeff. & -- \\
  Centralised Dimension Red. & $\mathbf{H}_l$, $MK$ coeff. & $\mathbf{A}_l$, $MN$ coeff. \\
  Decentralised Dimension Red. & $\mathbf{H}_l^{\scriptscriptstyle \mathrm{RD}}$, $NK$ coeff. & -- \\ \hline
\end{tabular}

\end{table}

\begin{figure}[h!]
\centering
\includegraphics[width=3.6in]{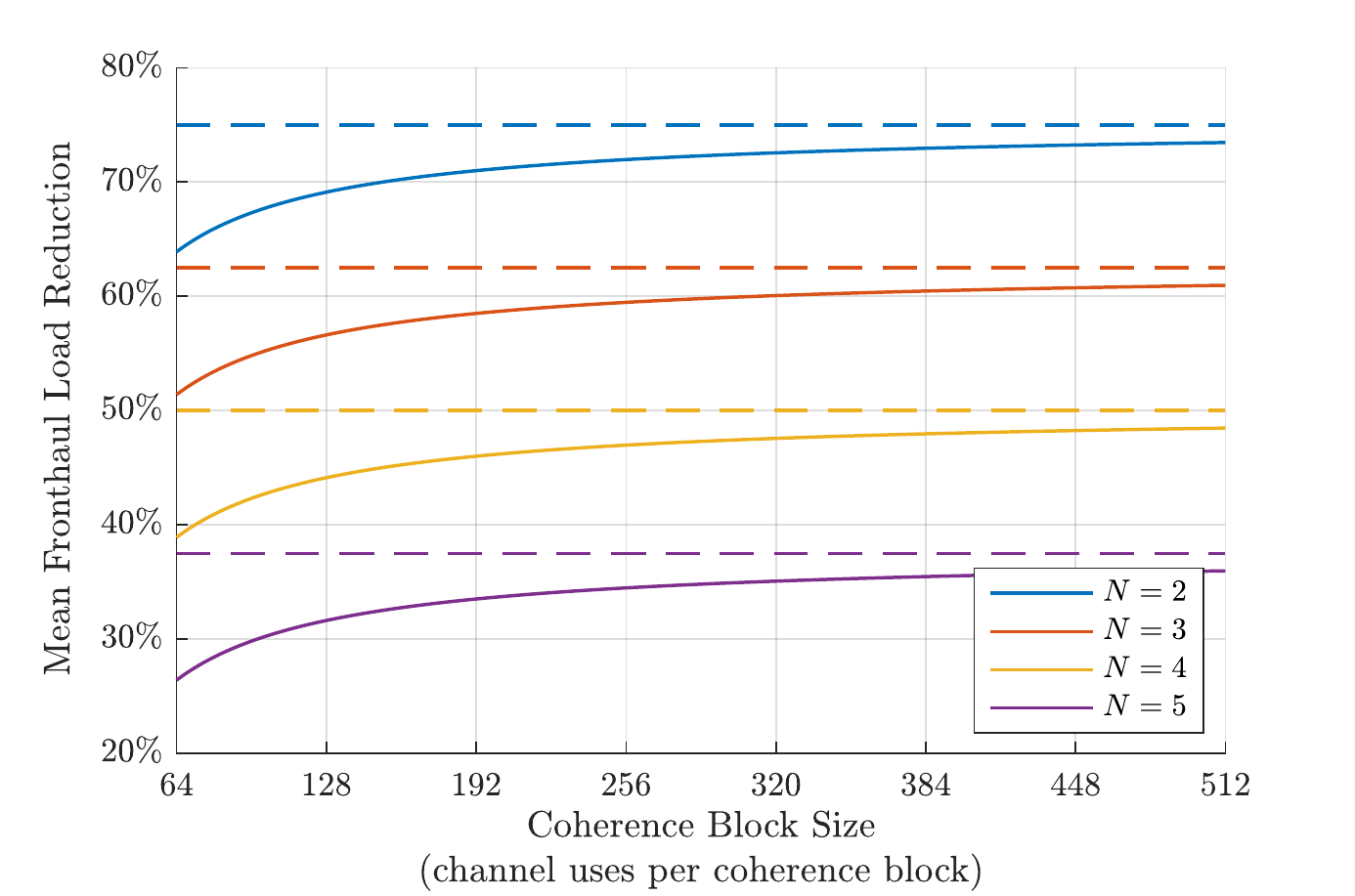}
\caption{Mean per-RRH fronthaul load reduction from dimension reduction (including CSI overheads), $K=8,L=4,M=8$. Solid line: centralised dimension reduction. Dot-dash line: decentralised dimension reduction.}
\label{fig:fronthaul_load}
\end{figure}

Figure \ref{fig:fronthaul_load} shows the reduction in total fronthaul load (including CSI overheads) in terms of mean coefficients per channel use, for a system with $K=8,M=8$. When the coherence block is small, CSI overheads account for a larger proportion of the fronthaul load, and hence the benefit from using decentralised dimension reduction is greater. 

\subsection{Imperfect CSI}
\label{sec:imperfectCSI}
In practical systems, the CSI estimated using uplink pilots is imperfect. Assuming the use of MMSE channel estimation, this can be accounted for by modelling the estimated channel between user $k$ and RRH $l$, $\hat{\mathbf{h}}_{l,k} \in \mathbb{C}^{M}$, as
\begin{equation}
	\hat{\mathbf{h}}_{l,k} = \mathbf{h}_{l,k} - \mathbf{e}_{l,k}
\end{equation}
where $\mathbf{h}_{l,k}$ is the propagation channel and $\mathbf{e}_{l,k}$ the random orthogonal channel error, with covariance $\mathbb{E}\{\mathbf{e}_{l,k}\mathbf{e}_{l,k}^\dagger\} = \mathbf{C}_{l,k}$. 

Following the reasoning in \cite{1193803}, the uplink received signal at RRH $l$ can be modelled as
\begin{align}
	\mathbf{y}_l &= \sqrt{\rho}\hat{\mathbf{H}}_l\mathbf{x} + \sqrt{\rho}\mathbf{E}_l\mathbf{x} + \bm{\eta}_l = \sqrt{\rho}\hat{\mathbf{H}}_l\mathbf{x} + \bm{\nu}_l
\end{align}
where $\bm{\nu}_l \in \mathbb{C}^M$ is a noise term that includes the effect of both receiver noise and orthogonal channel estimation error, with
\begin{equation}
\label{eq:CSI_cov}
	\mathbb{E}\{\bm{\nu}_l\bm{\nu}_l^\dagger\} = \mathbf{\Omega}_l = \mathbf{I}_M + \rho\sum_{k=1}^K	\mathbf{C}_{l,k}.
\end{equation}
We then apply a `whitening' transform to produce an equivalent MIMO system
\begin{align}
\label{eq:whitening}
	\check{\mathbf{y}}_l &= 	\mathbf{\Omega}_l^{-1/2}\mathbf{y}_l = \sqrt{\rho}\check{\mathbf{H}}_l\mathbf{x} + \check{\bm{\eta}}_l
\end{align}
where $\check{\mathbf{H}}_l = \mathbf{\Omega}_l^{-1/2}\hat{\mathbf{H}}_l$ and $\mathbb{E}\{\check{\bm{\eta}}_l\check{\bm{\eta}}_l^\dagger\} = \mathbf{I}_M$. The proposed dimension reduction methods may now be applied to $\check{\mathbf{y}}_l$, and, following the reasoning in \cite{1193803}, the uplink rates calculated using standard expressions and these equivalent channels represent \textit{expected} achievable rates under imperfect CSI. 

\squeezeup
\section{Downlink Dimension Reduction using Two-Stage Precoding}
\label{sec:DownlinkDimensionReduction}
\begin{figure}[!t]
\centering
\includegraphics[trim={6cm 1.5cm 21cm 4cm},clip,width=0.95\linewidth]{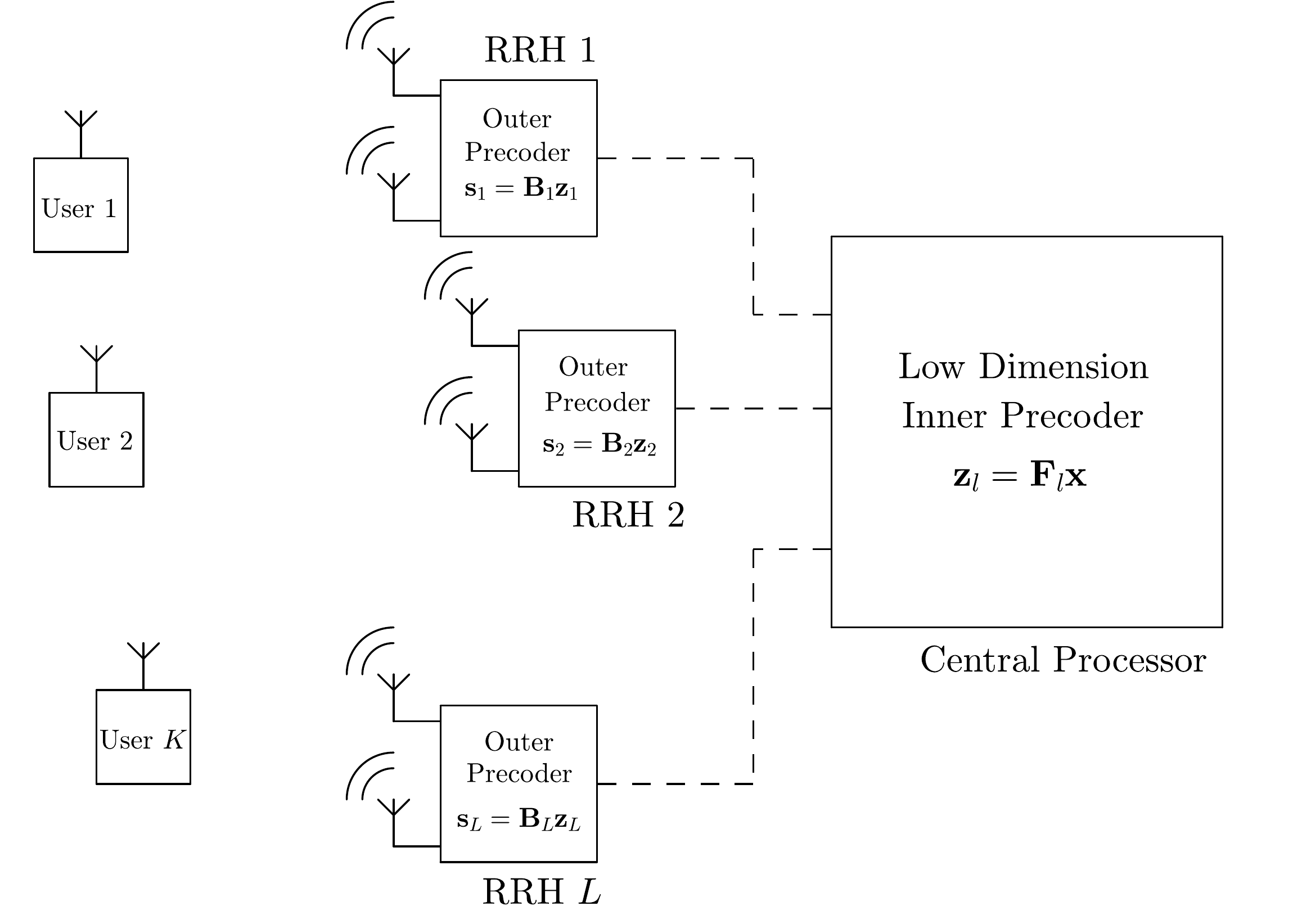}
\caption{Dimension reduction for downlink distributed massive MIMO C-RAN using two-stage precoding.}
\label{fig:Precoding}
\end{figure}

We now illustrate the potential for also applying dimension reduction on the distributed MIMO downlink, considering a `dual' of the uplink scheme that achieves dimension reduction using two precoding stages, as shown in Figure \ref{fig:Precoding}.

The inner precoding stage is applied to the downlink transmit symbols, $\mathbf{x} \sim \mathcal{CN}(0,\mathbf{I}_K)$, at the CP to produce $L$ low dimension signals, $\mathbf{z}_l \in \mathbb{C}^N$, which are then transferred over fronthaul to the RRHs. Here we restrict our attention to linear inner precoders, $\mathbf{F}_l \in \mathbb{C}^{N \times K}$, 
\begin{equation}
	\mathbf{z}_l = \mathbf{F}_l\mathbf{x}.
\end{equation}
The second, outer, linear precoding stage is then applied to the $N$-dimension signal at each RRH  to produce an $M$-dimension transmit signal,
\begin{equation}
	\mathbf{s}_l = \mathbf{B}_l\mathbf{z}_l.	
\end{equation}
The vector of received signals at the $K$ users is then
\begin{align}
	\mathbf{y} = \sum_{l=1}^L\bar{\mathbf{H}}_l^\dagger\mathbf{s}_l + \bm{\eta} = \sum_{l=1}^L\bar{\mathbf{H}}_l^\dagger\mathbf{B}_l\mathbf{F}_l\mathbf{x} + \bm{\eta}.
\end{align}
In principle $\mathbf{F}_l$ and $\mathbf{B}_l$ can be directly optimised according to some downlink performance criteria. We instead consider a simpler approach where the uplink dimension reduction filters are re-used in the outer precoding stage, i.e
\begin{equation}
	\mathbf{B}_l = \mathbf{A}_l.
\end{equation}
This is attractive from a practical perspective for TDD networks serving the same users on the uplink and downlink, as it removes the computation and signalling overheads associated with generating additional outer precoding matrices.

For semi-orthogonal $\mathbf{B}_l$, the per-RRH transmit power depends only on the inner precoders,
\begin{align}
\label{eq:powerconstraint}
	\mathbb{E}\{\mathbf{s}_l^\dagger\mathbf{s}_l\} &= \tr\big(\mathbf{F}_l^\dagger\mathbf{F}_l\big) \leq P,
\end{align}
These inner precoders can be thought of as acting on the set of reciprocal reduced dimension downlink channels, 
\begin{align}
	\bar{\mathbf{H}}_l^\dagger\mathbf{B}_l = \mathbf{P}^{-1/2}\mathbf{H}_l^{\scriptscriptstyle \mathrm{RD}\dagger},
\end{align}
and can be designed using standard downlink precoding methods, e.g. Moore-Penrose zero-forcing (ZF),
\begin{equation}
	\mathbf{F}_l = \mathbf{B}_l^\dagger\bar{\mathbf{H}}_l\Big(\sum_{j=1}^L\mathbf{B}_j^\dagger\bar{\mathbf{H}}_j\Big)^{-1}\mathbf{\Gamma}^{1/2},
\end{equation}
where $\mathbf{\Gamma} = \mathrm{diag}(\rho_k)$. Under perfect CSI, the downlink rate of user $k$ is then
\begin{equation}
	\mathcal{R}_k^{\scriptscriptstyle \mathrm{DL}}	= \log_2(1 + \rho_k),
\end{equation}
The $\rho_k$ can then be allocated to maximise some objective, such as max-min user rate, subject to the power constraints in \eqref{eq:powerconstraint}.

\squeezeup
\section{Numerical Results}
\label{sec:NumericalResults}

We now provide a selection of numerical results demonstrating the benefits of the proposed dimension reduction approach. These are obtained via Monte Carlo simulation for a dense urban environment, where the $K$ users and $L$ RRHs are distributed randomly within a 200 m $\times$ 200 m coverage area..  The user channels follow complex normal independent fading
\begin{equation}
	\mathbf{h}_{l,k} \sim \mathcal{CN}(0,\beta_{l,k}\mathbf{I}_M),	
\end{equation}
with a log-distance shadow fading path loss model,
\begin{equation}
	\beta_{l,k} = 147.5 - 20\log_{10}f_c -10\gamma\log_{10}d_{l,k} + \psi
\end{equation}
where $f_c$ is the carrier frequency in GHz, $\gamma$ the path loss exponent, $d_{l,k}$ the distance between RRH $l$ and user $k$ in metres, and $\psi \sim \mathcal{CN}(0,\sigma_\psi^2)$ the log-normal shadow fading term in dB. Simulation parameters are summarised in Table II, chosen based on the results given in \cite{7434656}.

\begin{table}
\centering
\label{tab:simconfig}
\caption{Simulation Configuration Parameters}
\begin{tabular}{c|c}
\hline
  Coverage area & 200 m $\times$ 200 m\\ 
\hline 
	User antenna height & 1 m\\ 
\hline 
	RRH antenna height & 6 m\\ 
\hline 
  Carrier frequency & $f_c = 3.5$ GHz\\ 
  \hline 
  Path loss exponent &  $\gamma = 2.7$ \\  \hline 
  Shadow fading variance & $\sigma_\psi^2 = 5.7$ \\ \hline
\end{tabular}
\end{table}

On the uplink, user power control is applied based on slow fading coefficients, such that each user has unity overall average power adjusted channel gain,
\begin{equation}
	 p_k\sum_{l=1}^L\frac{\beta_{l,k}}{L} = 1.
\end{equation}
All rates are averaged over both user \& RRH locations and random channel realisation. As a reference, the proposed schemes are compared to dimension reduction using joint antenna selection (adapted from \cite{1261322}), and simple antenna reduction (i.e. where each RRH has only $N$ antennas). 

\subsection{Uplink Sum Rate under MMSE-SIC Detection}
\begin{figure}[h!]
\centering
\includegraphics[width=3.6in]{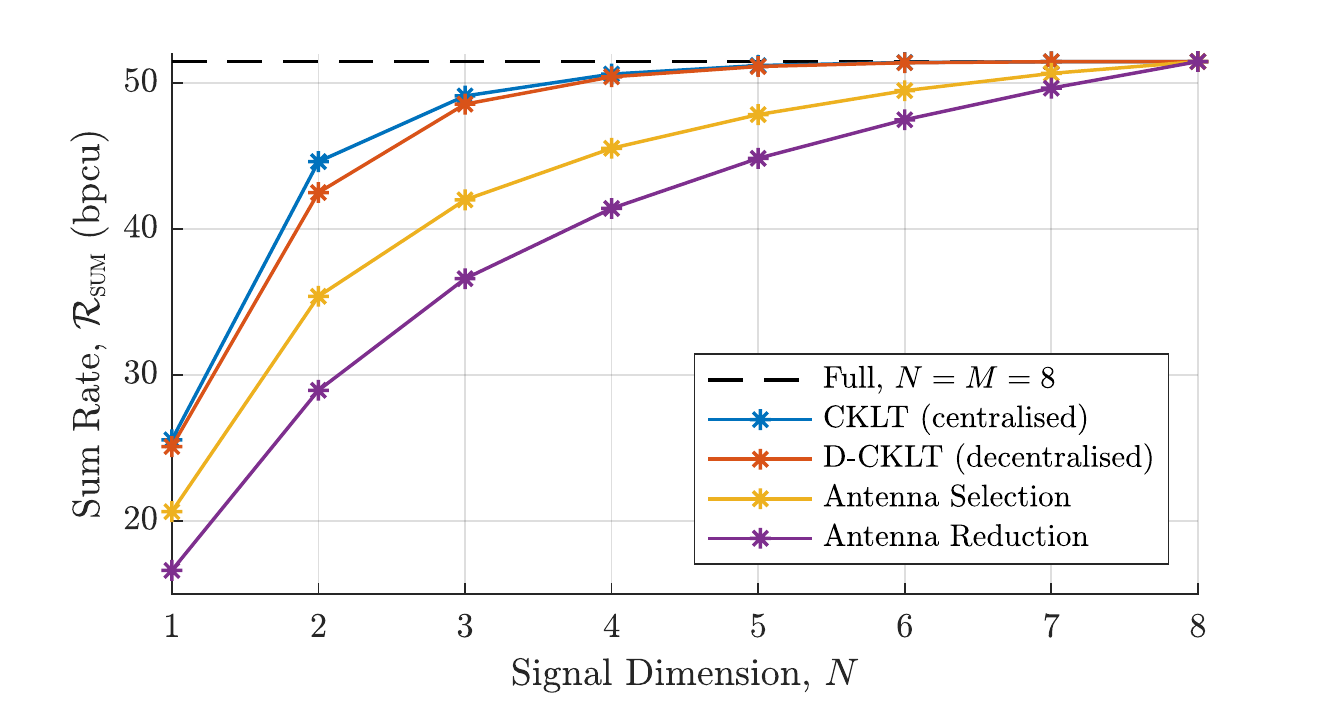}
\caption{Uplink sum rate under dimension reduction for varying signal dimension, $K=8,L=4,M=8,\rho = 5$ dB.}
\label{fig:MMSE_SIC_N}
\end{figure}

Figure \ref{fig:MMSE_SIC_N} shows the sum rates that can be achieved for varying reduced signal dimensions, $N$, in a system with $L=4$ RRHs, $K=8$ users, $M=8$ antennas and MMSE-SIC detection. At all reduced signal dimensions ($N<8$), the proposed centralised \& decentralised dimension reduction methods achieve significantly higher sum rates than can be achieved by either antenna selection or antenna reduction. 

For example, under either of the proposed schemes, reducing the signal dimension at each RRH from $N=M=8$ to $N=3$ results in only $\sim$5\% loss in cell sum rate, whilst reducing signal fronthaul data by 63\%. This compares favourably to simply reducing the number of antennas at each RRH to $M=3$, which incurs a $\sim$30\% loss in sum rate.

\begin{figure}[h!]
\centering
\includegraphics[width=3.6in]{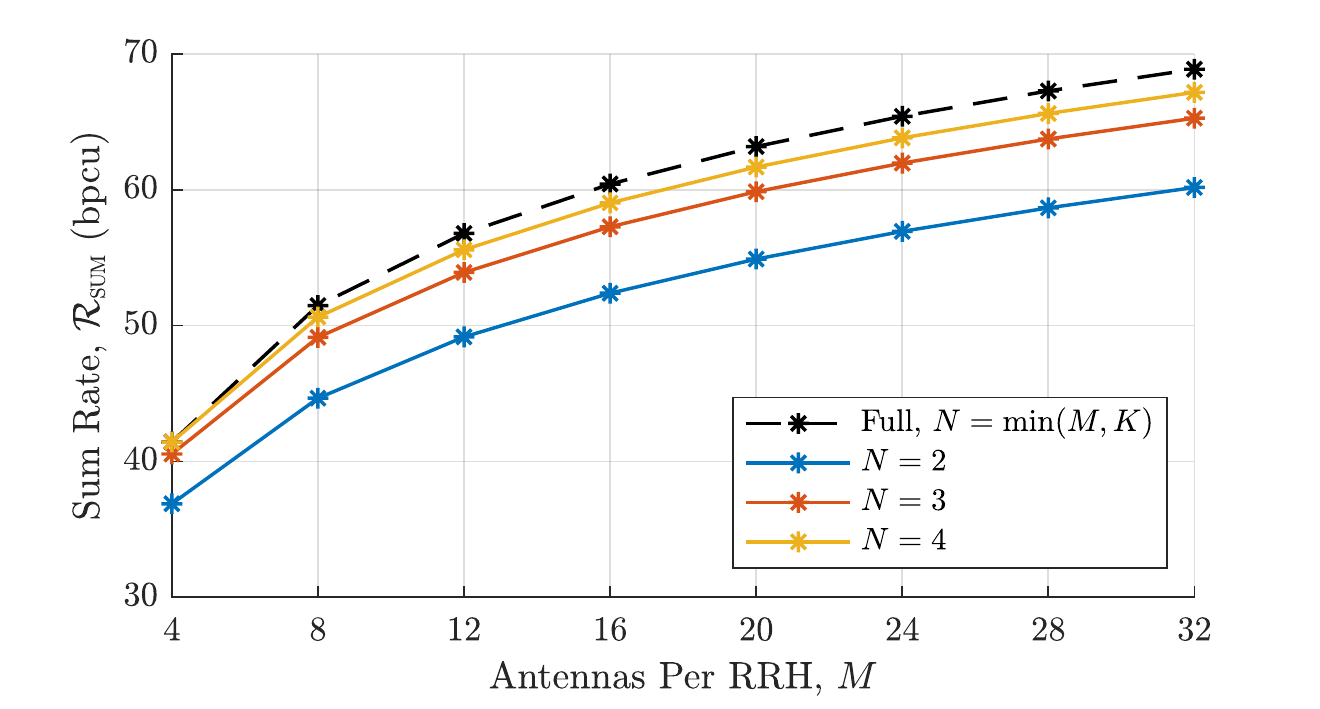}
\caption{Uplink sum rate under CKLT dimension reduction for varying number of antennas at each RRH, $K=8,L=4,\rho = 5$ dB.}
\label{fig:MMSE_SIC_M}
\end{figure}

Figure \ref{fig:MMSE_SIC_M} shows the case where the signal dimension at each RRH, $N$, is kept constant but the number of antennas deployed at each RRH, $M$, is increased, with centralised (CKLT) dimension reduction. Despite the fixed dimension signal, the cell sum rate is always increased by increasing the number of antennas -- demonstrating the benefit of operating in the `massive' regime even when fronthaul capacity is limited. 

\subsection{Uplink User Rates under Linear MMSE Detection}
\begin{figure}[h!]
\centering
\includegraphics[width=3.6in]{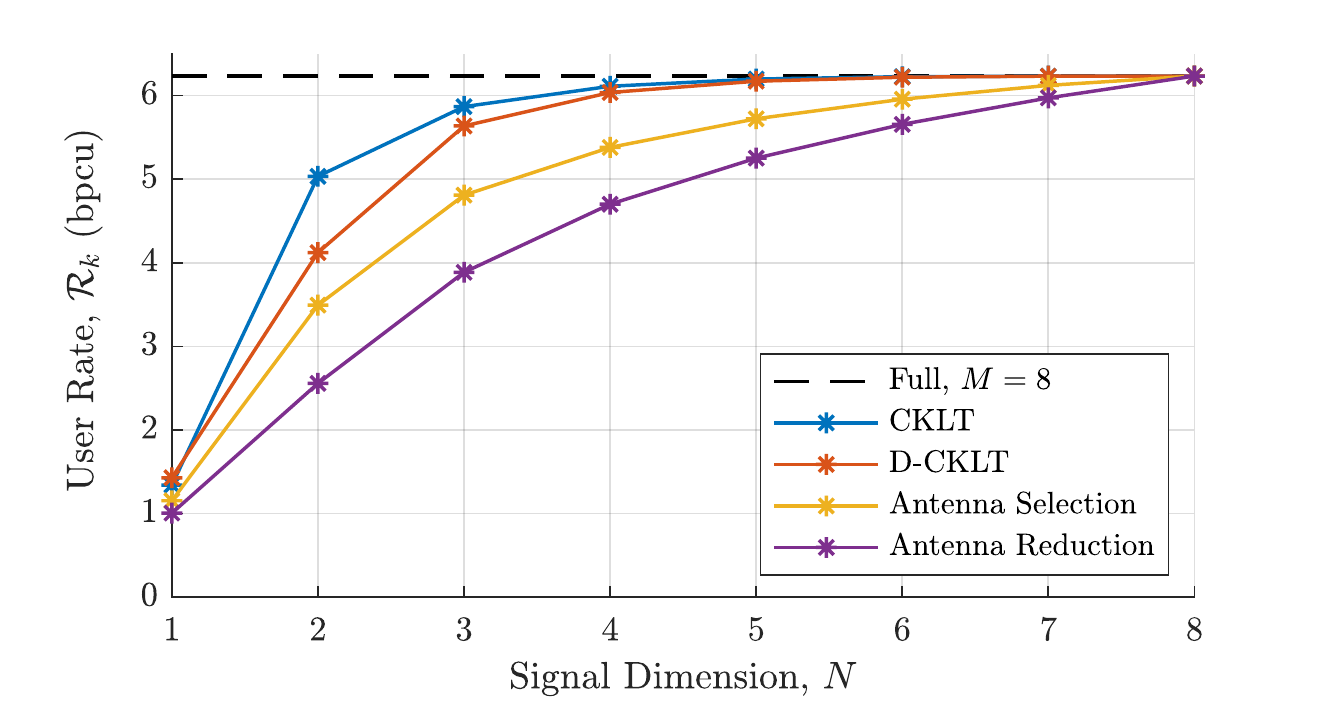}
\caption{Uplink user rates under dimension reduction with linear MMSE detection, $K=8,L=4,M=8,\rho = 5$ dB.}
\label{fig:MMSE_N}
\end{figure}

Figure \ref{fig:MMSE_N} shows the mean user rates under linear MMSE detection for the two proposed dimension reduction techniques, $K=8,L=4,M=8$. Note that for $N=1$ the overall equivalent reduced dimension MIMO system is degraded (rank deficient), and linear detection cannot effectively isolate the user streams, resulting in poor performance. For $N > 2$ the reduced dimension a high proportion of the full user rate is achieved.

\subsection{Uplink User Outage Performance}
\begin{figure}[h!]
\centering
\includegraphics[width=3.6in]{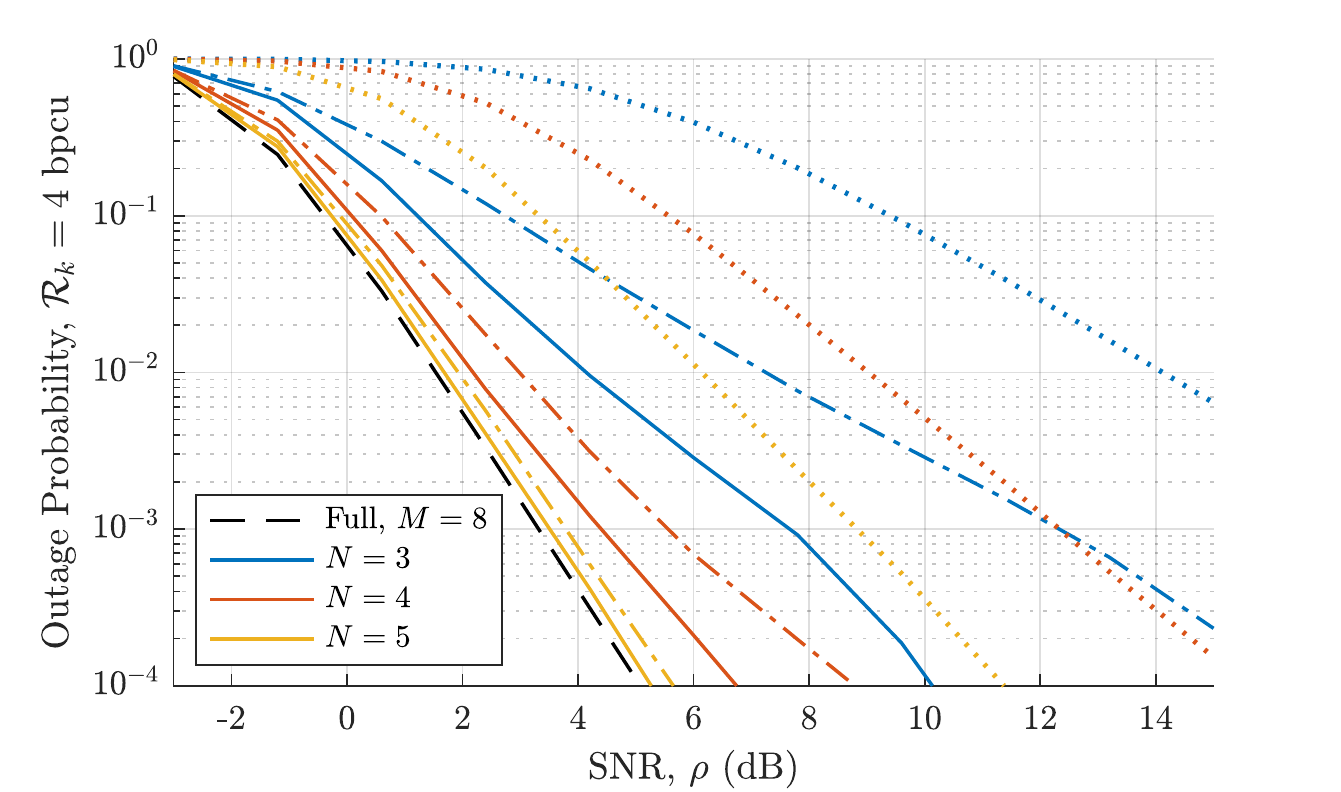}
\caption{Uplink user outage probability under MMSE detection, $\mathcal{R}_k = 4$ bpcu, $K=8,L=4, M=8$. Solid line: CKLT, Dot-dash line: D-CKLT, dotted line: antenna reduction ($M'=N$).}
\label{fig:outage}
\end{figure}

Figure \ref{fig:outage} shows the user outage performance under linear MMSE detection for varying numbers of antennas (top), and fixed number of antennas ($M=8$) with dimension reduction applied. The benefits of deploying excess antennas and then performing dimension reduction to reduce fronthaul load are again clear for both centralised and decentralised cases. For $N=5$, the outage performance is close to that of the full dimension system. A theoretical analysis of the outage performance remains as an interesting topic for future work.

\subsection{Scalability}
\begin{figure}[h!]
\centering
\includegraphics[width=3.6in]{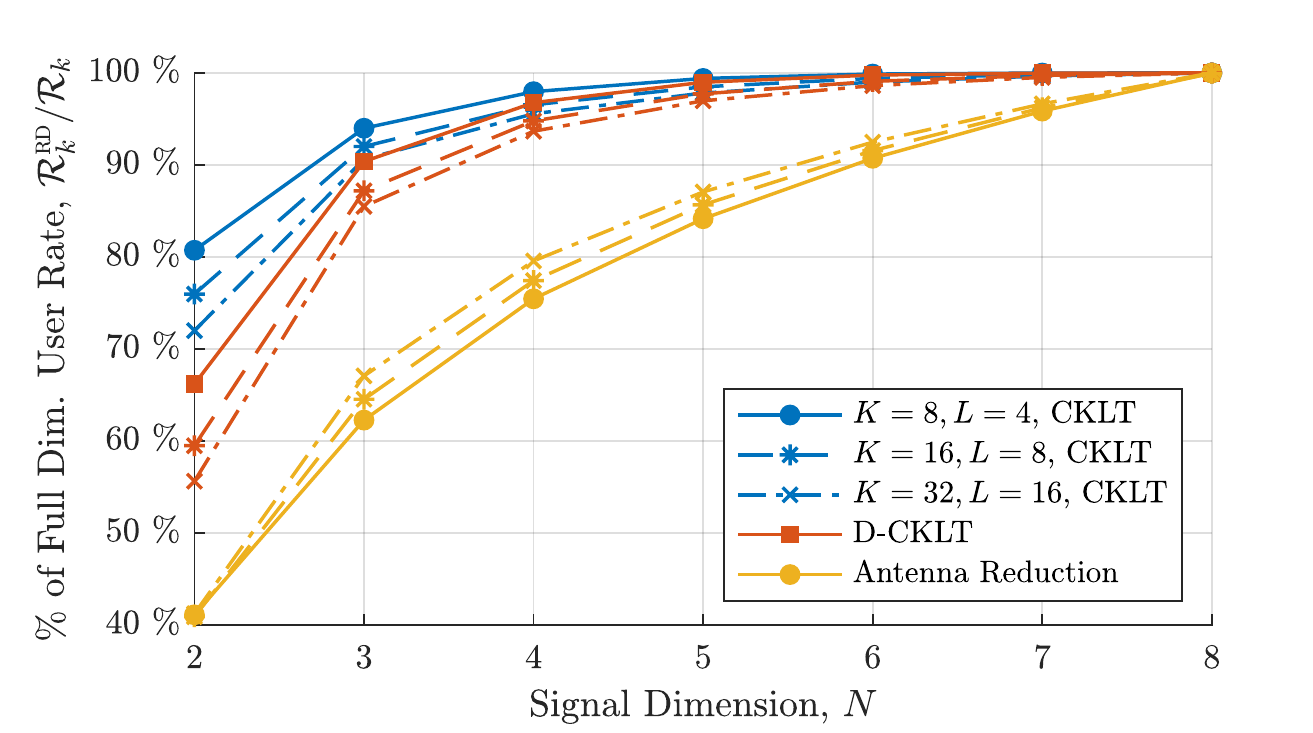}
\caption{Uplink user rate under dimension reduction compared to full dimension MIMO system as network density increases, $M=8, \rho = 5$ dB. Solid line: $K=8, L=4$. Dashed line: $K=16, L=8$, Dot-dash line: $K=32, L= 16$.}
\label{fig:scaling}
\end{figure}

Figure \ref{fig:scaling} shows that as the network becomes denser -- by adding more users and RRHs -- dimension reduction continues to capture a very high proportion of the full dimension MIMO system throughput. We note that whilst there is some performance penalty compared to the centralised approach, the fully decentralised approach (D-CKLT) has a much reduced computational load that is distributed between the RRHs, and therefore represents an attractive scalable solution for use in large C-RAN networks. 
\squeezeup
\subsection{Imperfect CSI}
\begin{figure}[h!]
\centering
\includegraphics[width=3.6in]{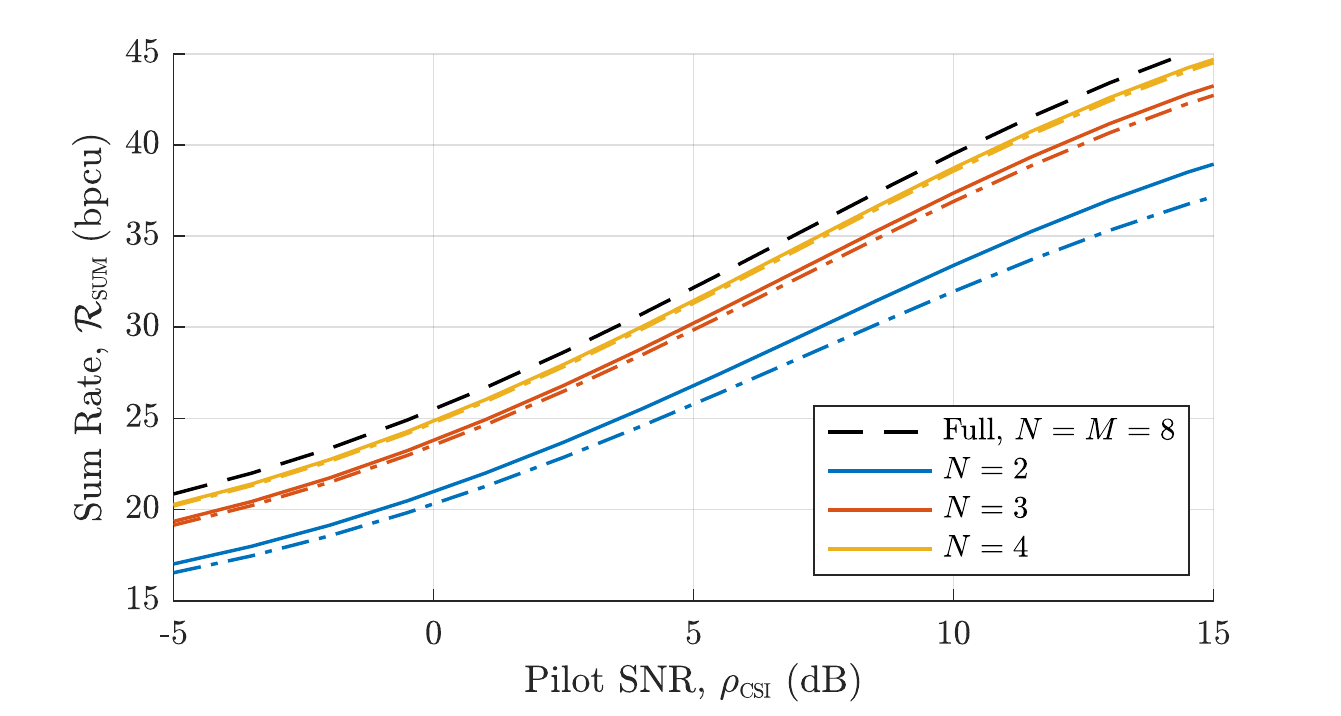}
\caption{Achievable uplink sum rate for varying CSI quality under centralised dimension reduction, $K=8,L=4,M=8, \rho=5$ dB. Solid line: CKLT. Dot-dash line: D-CKLT.}
\label{fig:CSI}
\end{figure}

We now consider the case of imperfect (noisy) CSI at the RRHs. Figure \ref{fig:CSI} shows the achievable sum rate, based on the model discussed in Section \ref{sec:imperfectCSI}, assuming the MIMO channels are estimated from orthogonal pilot signals with SNR $\rho_{\scriptscriptstyle \mathrm{CSI}}$. The performance of the reduced dimension system is close to that of the full dimension system irrespective of the CSI quality.

\subsection{Downlink Two-Stage Precoding}
\begin{figure}[h!]
\centering
\includegraphics[width=3.6in]{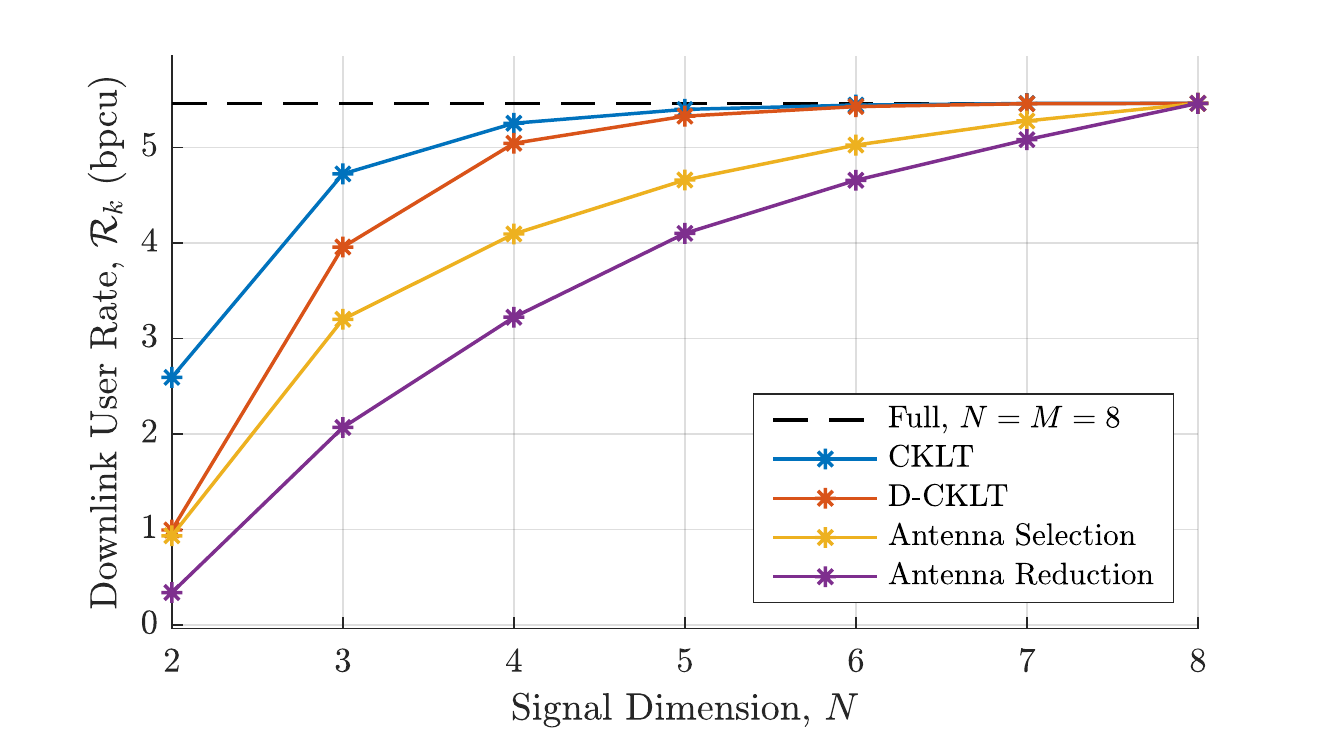}
\caption{Downlink mean user rates for two-stage precoding with max-min user power control, $K=8,L=4,M=8,P=9$ dBm.}
\label{fig:DL}
\end{figure}

As discussed in Section \ref{sec:DownlinkDimensionReduction}, the dimension reduction approach to fronthaul data reduction readily extends to the downlink in the form of a two-stage precoding scheme. Figure \ref{fig:DL} illustrates that similar performance benefits are afforded as on the uplink -- establishing dimension reduction as an effective bidirectional approach to fronthaul data reduction for massive MIMO systems. 

\section{Conclusion}
This paper has provided an investigation into the use of dimension reduction for reducing fronthaul data in massive MIMO C-RAN systems. Centralised and decentralised methods for designing appropriate dimension reduction filters have been proposed, and a combination of analysis and numerical results used to show that dimension reduction, applied locally at each RRH, can significantly reduce fronthaul traffic without incurring significant loss in cell throughput. Furthermore, we have shown that this dimension reduction is readily extended to the downlink of MIMO systems, establishing it as an effective and practical bidirectional approach to fronthaul data reduction that is particularly well suited to TDD networks. 

Overall, our work shows that it is possible to use dimension reduction to produce an equivalent reduced dimension distributed MIMO system that preserves the performance benefits of operating in the `massive' regime with a large overall excess of BS antennas. Future work should build on these contributions to further analyse the behaviour and performance of reduced dimension MIMO systems, and investigate \& implement dimension reduction schemes in practical systems.

\section*{Acknowledgment}
The authors would like to thank Toshiba Europe Ltd. and EPSRC for financially supporting this work.
\IEEEtriggeratref{4}

\ifCLASSOPTIONcaptionsoff
  \newpage
\fi



\bibliographystyle{IEEEtran}
%

\bibliographystyle{IEEEtran}
\bibliography{TVTbib}

\end{document}